%% file: main.tex
\documentclass[11pt]{article}
\usepackage{amsmath,amssymb,amsfonts}
\usepackage{latexsym}
\usepackage{pstricks}  

\usepackage{natbib}

\newcommand{\beqn}{\begin{eqnarray}\begin{aligned}}
\newcommand{\eqn}{\end{aligned}\end{eqnarray}}

\usepackage{graphicx}

\begin{document}
\begin{titlepage}
\begin{flushright}
UTAS-PHYS-2009-\hspace*{.2cm}\\
September 2009\\
\end{flushright}
\begin{centering}
 
{\ }\vspace{0.5cm}
 
{\Large\bf Markov invariants for phylogenetic rate matrices }

\vspace{5pt}

{\Large\bf   derived from embedded submodels}

\vspace{1.8cm}

P. D. Jarvis\footnote{Alexander von Humboldt Fellow}${}^,$\footnote{Tasmanian Fulbright Scholar 2009} and
J. G. Sumner\footnote{Australian Postdoctoral Fellow}

\vspace{0.3cm}

{\em School of Mathematics and Physics}\\
{\em University of Tasmania, Private Bag 37}\\
{\em Hobart, Tasmania 7001, Australia }\\
{\em E-mail: {\tt Peter.Jarvis@utas.edu.au}, {\tt Jeremy.Sumner@utas.edu.au}}

\thispagestyle{empty}
\begin{abstract}
\noindent
We consider novel phylogenetic models with rate matrices that arise 
via the embedding of a progenitor model on a small number of character states, 
into a target model on a larger number of character states.
Adapting representation-theoretic results from recent investigations of 
Markov invariants for the general rate matrix model, we give a 
prescription for identifying and counting Markov invariants for such 
 `symmetric embedded' models, and we provide enumerations of these for low-dimensional cases.
The simplest example is a target model on 3 states, constructed 
from a general 2 state model; the 
`2$\hookrightarrow$3' embedding. We show that for 2 taxa, there exist two invariants of 
\emph{quadratic} degree, that can be used to directly infer pairwise distances
from observed sequences under this model.
A simple simulation study verifies their theoretical expected values, 
and suggests that, given the appropriateness of the model class, they have greater statistical power than the standard (log) Det invariant (which is of \emph{cubic} degree for this case).

\end{abstract}

\noindent

\vspace{10pt}

\end{centering} 

\vspace{125pt}

\end{titlepage}

\setcounter{footnote}{0}

\input{intro}
\input{sec1}
\input{sec2}
\input{sec4}

\input{conc}

\input{appendix}

\subsection*{Acknowledgements}
This research was supported by Australian Research Council Discovery Grant DP0877447.  

\bibliographystyle{jtbnew}
\bibliography{masterAB}

\end{document}

%% file: intro.tex
\section{Introduction}
\label{sec:Introduction}
Phylogenetic inference based on molecular sequence data typically involves the selection of one or more specific models for state substitutions. 
There is a well-known hierarchy of classes of $4 \times 4$ rate matrices, with varying complexity and numbers of free parameters \citep{posada1998}.
However, for a given data set it is not always (if ever) clear which substitution model is most appropriate or ``best''.  
For example, the software package \texttt{ModelTest} \citep{posada1998} selects a model of nucleotide substitution that best fits a given data set under a likelihood, information theoretic or Bayesian framework.
While it is useful to be able compare results derived from different models, given the dangers of over-parametrisation the question of which datasets best conform to which class of models is difficult to resolve. 
Elaborations such as allowing rate variation across sites and invariant sites are also standard ingredients which allow for more flexibility in data-fitting.
More extreme measures, which relinquish the conventional picture of species evolution via Markov models on trees, lead to generalisations such as mixtures \citep{pagel2004} (sites which have  probabilities for following different models), ``mosaics'' \citep{woodhams2009} (edge classes or subtrees which have different weights for different models), or ultimately network models \citep{bandelt1992,holland2004}, for example using acyclic directed graphs.

In recent work we have introduced so-called ``Markov invariants'' \citep{sumner2008,sumner2009}, which are polynomial quantities built up out of the phylogenetic pattern frequencies or divergence arrays.
Markov invariants are distinct from ``phylogenetic invariants'' \citep{cavender1987,lake1987} in that they are defined to behave as a (one-dimensional) ``representation'' of continuous unfolding of the Markov process.
This means that as the Markov process proceeds in time by an amount $\tau$, the expectation value of the Markov invariants simply scales with a product of the multiplicative constants $\det(m_e)=e^{tr(Q_e)\tau}$, where $m_e=e^{Q_e\tau}$ is the transition matrix associated with the edge $e$.  
With this understanding, the ``invariance'' property of the Markov invariants is captured by the simple time-dependence
\beqn
\tau \rightarrow \tau+\tau' \quad\Longleftrightarrow\quad  e^{tr(Q_e)\tau}\rightarrow e^{tr(Q_e)\tau}e^{tr(Q_e)\tau'}=e^{tr(Q_e)(\tau+\tau')}.\nonumber
\eqn
It is important to note that the definition of phylogenetic invariants stipulates no such constraint.  

Markov invariants are phylogenetically informative for the most general phylogenetic model, giving some information of both model parameters \textit{and} tree topology,  and can be implemented \emph{without} the need for explicit parameter estimation (ie. via optimization of a likelihood function). 
These invariants generalise the ``log Det" distance measure which has precisely this feature: pairwise distances can be directly estimated whose expected value turns out to be the sums of rate parameters multiplied by time. 
We have identified Markov invariants for diverse combinations of numbers of taxa and numbers of characters. 
For example there are three so-called ``squangle'' invariants \citep{sumner2008,sumner2009} for quartets of taxa and four character states, whose values directly resolve and distinguish the three unrooted trees $12|34$, $13|24$, $14|23$ for the most general Markov model, without the need for parameter estimation. 
Indeed, there are consistency arguments to suggest that the Markov invariants, as defined simply as functions of the phylogenetic pattern data, are in fact identical to their maximum likelihood estimators (or, more technically, belong to the ideal generated by the solutions of the likelihood equations).
This is trivially true for the log Det estimator \citep{allman2009a} and its proof in the general case is a subject of future work.

Notwithstanding these promising developments for the general Markov model, it is still of great interest to have at hand tools for exploring the full range of models available for phylogenetic inference. In this vein, \citet{huelsenbeck2004} identified up to 203 submodels of the general time reversible model (GTR) for four characters, the number being based simply on the combinatorial problem of counting compositions and refinements as parameters are turned on and off at various positions in the rate matrix.

In view of the discussion above concerning model classes, a natural criterion for model and sub-model selection is that of the multiplicative closure of the edge transition matrices, or semigroup property. 
Multiplicative closure is sometimes held out to be required for establishing the unrootedness of phylogenetic trees \citep{isaev2004,semple2003}, but for a given tree it seems not to be necessary. 
Its importance arises more directly from the methodology and interpretation of phylogenetic reconstruction. 
In doing tree searches for example, a swing of a leaf edge from one subtree to another entails a cut-and-rejoin operation: the incoming and outgoing Markov edge matrices from the source node of the originating leaf edge must be multiplied, while the Markov matrix from the target edge where the leaf is rejoined, must be expressed in turn as the product of Markov matrices for two new edges.
Again, the possibility of extinctions along some edges, or of incomplete taxon sampling, suggests that, to allow for correct marginalisation, multiplicativity is mandatory for a consistent interpretation. 
It is clear that scarcely any of the GTR matrices identified by \citet{huelsenbeck2004} will be multiplicative (indeed, only some well-known models comply, for example the symmetric models such as the Kimura model, as well as Felsenstein's TR, non-symmetrical model).
Indeed it is easy to show that the GTR model itself is not multiplicative, and this poses serious interpretive questions if the GTR class is used in generalized models where more than a single rate matrix is implemented in the analysis.

In this paper we introduce, and explore through Markov invariants, new classes of submodels for a given number of characters, generated from general models in smaller numbers of characters. 
These we term `symmetric embeddings'. 
Clearly, in essence, these submodels contain similar information to the originating model in lower dimensions (with a smaller number of characters, and fewer parameters), and it is the manner in which this intuition is realised in technical detail, which we wish to elaborate here. 
In contrast to the na\"{i}ve identification of submodels by the mere presence or absence of additional parameters however, our embedded models are by construction multiplicatively closed. 
Moreover we are able to adapt the technical setting of Markov invariants to this new situation, and so derive new invariants of different structure and polynomial degree from the standard ones, which play an equivalent role to them. 
Again, these new invariants fulfill the expectation (because their underlying models have fewer parameters) that they should be of lower degree than the standard ones.

In \S \ref{sec:SymmEmbedd} below we introduce the symmetrically embedded models. 
These are given in a general setting, but we concentrate in detail on the general 2 state model embedded into 3 character models, called the `$2 \hookrightarrow 3$' case. 
A variety of Markov invariants, for diverse degrees and numbers of characters and taxa, is enumerated in \S \ref{sec:MarkovInv} after adapting the group representation method for identifying Markov invariants \citep{sumner2008,sumner2009} to the present setting. 
The simplest case is again that of `$2 \hookrightarrow 3$', with two taxa, where it is shown that, apart from the (degree 3) determinant  function (guaranteed to exist for the general Markov model and any number of characters, and well known as the log Det measure), there are two additional quadratic degree invariants called $I_{3,1}$ and $I_{2,2}$. 
These are constructed explicitly and their properties are explored.

 
Finally in \S \ref{sec:Results} the paper is summarised, and the conclusions supported by some simple simulated data analysed for comparison using  ``Det'' invariant, or $I_{3,3}$ in our notation, as well as $I_{3,1}$ and $I_{2,2}$ invariants. 
As expected, the invariants of lower degree (and ``weight'', see below), are apparently statistically better behaved, at least from this preliminary numerical test. 
The paper ends with some concluding remarks and prospects for further work.  
An appendix, \S \ref{sec:appendix}, gives an adaptation of a technical representation-theoretic result from \citep{sumner2008} enabling the Markov invariants for embedded submodels to be enumerated and constructed as presented in \S 3 for low-dimensional cases.

%% file: sec1.tex
\section{Symmetrically embedded character substitution models}
\label{sec:SymmEmbedd}
In this section we introduce the concept of symmetrically embedded models, concentrating initially on the two state case, and developing the analysis to be able to present the $2\hookrightarrow 3$, $2\hookrightarrow 4$ and $2\hookrightarrow K$ rate matrices in detail.
The discussion finishes with an overview of the general case.

Consider the rate matrix for the general Markov model on two characters, 
\begin{align}
\label{eq:Qmatrix2by2}
 Q  =& \, \left( \begin{array}{cc}\!-\!\alpha&\beta \\ \alpha & \!-\!\beta \end{array}\right) =
 \alpha \left( \begin{array}{cc} \!-\!1&0 \\ 1 & 0 \end{array}\right) + \beta \left( \begin{array}{cc} 0&1 \\ 0 & \!-\!1 \end{array}\right) \equiv \alpha L_\alpha + \beta L_\beta
 \end{align}
where as usual in the Markov matrix $m(t):= \exp(Qt)$, the matrix elements $m_{ab}$ have the interpretation 
\[
m_{ab} = {\mathbb P}\left[X(t)=a \mid X(0)=b\right]
\]
for the random variable $X(t)$ in character space\footnote{In our notation the the random change process is implemented by the left matrix action, so that the column sum of $m$ is unity (the column sum of $Q$ vanishes).}  describing the probability of change (along each edge after time $t$),
\begin{align}
\label{eq:LinearAction}
p_a(t) = \sum_{b=1}^2 m_{ab}p_b(0) ,
\end{align}
or $p(t) = m(t)\cdot p(0)$, where we have $K=2$ characters and the edge probability distribution is 
$p_a(t) = {\mathbb P}(X(t)=a)$.

We have chosen to write $Q$ in terms of the natural basis of column-sum zero `stochastic generator matrices' $\{L_\alpha,L_\beta\}$ of the group $GL_1(2)$; the subgroup of the general linear group $GL(2)$ of invertible $2\times 2$ matrices together with the probabilitistic constraint of unit-column sums \citep{johnson1985,mourad2004}. 
This is relevant for considerations of multiplicative closure of models, which might arise in applications where different rate matrices are allowed on different parts of a phylogenetic tree; where potentially missing taxa may need to be inserted into edges; or where re-evaluations of phylogeny may require edge rearrangements. 
In this case of a \emph{general} Markov rate model, in continuous time, closure of the product $M_1M_2 = \exp Q_1 \exp Q_2$ is guaranteed by the so-called BCH formula which requires closure of the \emph{commutator} brackets ${[} Q_1, Q_2 {]} := Q_1Q_2-Q_2Q_1 $ of the $Q$'s:
\beqn\label{eq:bch}
\exp Q_1 \exp Q_2 = \exp ( Q_1+Q_2 +\frac 12 {[}Q_1, Q_2 {]} + \frac{1}{12} {[}Q_1, {[}Q_1, Q_2 {]}{]} - \frac{1}{12} {[}Q_2, {[}Q_1, Q_2 {]}{]} + \cdots ).
\eqn


Referring to the BCH formula (\ref{eq:bch}), it is immediately clear that closure is assured if the rate matrices form a \emph{Lie algebra}, which in this case follows as we have chosen the most general two-state model. 
Specifically,
\[
{[} L_\alpha, L_\beta {]} = L_\alpha - L_\beta=-{[} L_\beta, L_\alpha {]} 
\]
with of course ${[} L_\alpha, L_\alpha {]} = 0 = {[} L_\beta, L_\beta {]} $.

How can we use (\ref{eq:Qmatrix2by2}) to infer rate matrices for `target' models on different  numbers of characters (larger than 2)? 
A natural observation from the linear $m$-action on the array (vector) $p_a$ is that a similar linear action can be obtained not only on the components of $p$, but also on any homogeneous polynomials in the components. 
Specifically we can regard the $k+1$ distinct monomials in components of the $p$ at fixed degree $k$, as the formal components of a new, $k+1$-dimensional array.

Consider for instance the case $k=2$, $k+1=3$, and the monomials $p_1^2$, $p_1p_2=p_2p_1$, and $p_2^2$. We write \emph{four} terms to emphasize that the new character probabilities, say $P_{1}=p_1^2$, $P_{2}=2 p_1 p_2$, $P_{3}=p_2^2$ are really components of a \emph{symmetric} array (tensor) $p_{ab}:= p_a p_b$, with of course $p_{ab}=p_{ba}$, and to motivate the choice of scaling; we use the bionomial expansion $(p_1+p_2)^2=p_1^2+2p_1p_2+p_2^2=1$.
If we take the differential form of the change rule (\ref{eq:LinearAction}), $dp/dt = Qp$, and write the induced transformation on $P$ (considered as a three component vector) as $dP/dt=Q^{(3)}P$, it is then easy to infer $Q^{(3)}$ by considering $dp_{ab}/dt$ and referring to (\ref{eq:Qmatrix2by2}). 
Following this through we find that
\begin{align}
\label{eq:Qmatrix3by3}
 Q^{(3)}  =& \,  \left( \begin{array}{ccc} \!-\!2 \alpha&\beta&0 \\ 2\alpha & \!-\! \alpha\!-\!\beta & 2\beta
\\ 0 & \alpha &\!-\!2 \beta  \end{array}\right) =
 \alpha \left( \begin{array}{ccc} \!-\!2&0 &0\\ 2 & \!-\!1&0 \\0&1&0 \end{array}\right) +
 \beta \left( \begin{array}{ccc} 0&1 &0\\ 0& \!-\!1&2 \\0&0&\!-\!2 \end{array}\right)
 \equiv \alpha L^{(3)}_\alpha + \beta L^{(3)}_\beta .
 \end{align}
By construction, the new generator matrices $L^{(3)}_\alpha$ and $L^{(3)}_\beta$ form a subalgebra of the Lie algebra of $GL_1(3)$ and satisfy the \emph{same} commutation relations as their $2\times 2$ progenitors, namely ${[}L^{(3)}_\alpha, L^{(3)}_\beta{]} = L^{(3)}_\alpha-L^{(3)}_\beta$. 
Thus technically we have an \emph{embedding} of the Lie algebra of $GL_1(2)$ into that of $GL_1(3)$, which we shall denote $2 \hookrightarrow 3$ (and of course multiplicative closure for this class of $3 \times 3$ model is guaranteed). 

The generalisation to the $2 \hookrightarrow 4$, or $2 \hookrightarrow K$, character case is immediate. For 
$K=4$ we have
\begin{align}
 Q^{(4)}  = \,  \left( \begin{array}{cccc} \!-\!3 \alpha\hskip-1ex&\beta&0&0 \\ 3\alpha & \!-\! 2\alpha\!-\!\beta \hskip-1ex& 2\beta&0
\\ 0 &2\alpha& \!-\!\alpha\!-\!2\beta \hskip-1ex &3 \beta \\ 0&0&\alpha&\!-\!3\beta \end{array}\right) &=
 \alpha \left( \begin{array}{cccc} \!-\!3  &0&0&0 \\ 3  & \!-\! 2  & 0&0
\\ 0 &2 & \!-\!1 &0 \\ 0&0&1&0 \end{array}\right) +
 \beta \left( \begin{array}{cccc}0&1&0&0 \\  0 & \!-\!1 & 2 &0
\\ 0 &0& \!-\!2 &3  \\ 0&0&0&\!-\!3  \end{array}\right)\nonumber\\
 &\equiv  \alpha L^{(4)}_\alpha \!+\!\beta L^{(4)}_\beta ,
 \nonumber
 \end{align}
based on a totally symmetric, rank three tensor $p_{abc}:= p_ap_bp_c$ with binomial constants of proportionality derived, as above, and using the constraint $(p_1+p_2)^3=1$ to form the vector $P$ with four components $P_{1}=p_1^3$, $P_{2}=3p_1p_2^2$, $P_{3}=3p_1^2p_2$ and $P_{4}=p_2^3$. 
In the $2 \hookrightarrow K$ case (corresponding to a rank $k=K\!-\!1$ tensor array $p_{a_1 \cdots a_{k}}$), the rate generator matrices  $L^{(K)}_\alpha$, $L^{(K)}_\beta$ have lower and upper diagonal entries $K\!-\!1, \cdots, 2,1$ and $1, 2, \cdots, K\!-\!1$, respectively, with the diagonals ensuring that the zero column sum condition is satisfied.  
Again these matrices satisfy the same commutator bracket relations (Lie algebra) as their $2\times 2$ progenitors and hence generate phylogenetic models that are guaranteed to satisfy the closure property.

%% file: sec2.tex
\section{Markov invariants}
\label{sec:MarkovInv}

\noindent
In this section we adopt the background context of group actions on which we base our general theorems on Markov invariants. Details are provided in \S \ref{sec:appendix}, where we restate our previous technical results. We then explore the counting of Markov invariants for symmetrically embedded models in diverse dimensions (number of characters of the generating model, embedding rank and hence number of characters of the derived model, number of taxa, and polynomial degree of the invariant) and tabulate several low-dimensional cases.
Finally we give details of the quadratic degree invariants for the $2\hookrightarrow 3$ case, and explicit constructions of them along with the cubic, determinant function for comparison.

As explained above and in systematic terms in \S \ref{sec:appendix}, embedded submodels are associated with particular matrix group constructions, whereby the character probability distribution $p$ for a starting model on ${K'}$ characters, is regarded as a progenitor for a target model deriving from a composite tensor array. If the starting model has dimension $K'$ and the tensor $p$ is of rank $k$ and totally symmetric (the only case we consider), then $p$ has $K= K'(K'\!+\!1)(K'\!+\!2)\ldots(K'\!+\!k\!-\!1)/k!$ components, and so $K$ is number of characters of the the target model. Table \ref{tab:ModelExx} gives a list of several cases of interest for low-dimensional examples; of course $K'=2$, $k=2,3, \cdots K\!-\!1$ are the $2\hookrightarrow 3$, $2\hookrightarrow 4$, and $2\hookrightarrow {K}$ cases already identified. 

\begin{table}[tbp]
  \centering 
  \begin{tabular}[tbp]{|c||c|c|c||c|c||c|c|}
\hline
$K'$ & 2&2&2&3&3&4&4 \\
\hline
$k$ & 2&3&4&2&3&2&3 \\
\hline
$K$ & 3&4&5&6&10&10&20 \\
\hline
\end{tabular}
  \caption{Identification of embedded submodels for low-dimensional cases of interest. $(K',k,K)$ gives the number of characters of the progenitor model, the rank of the embedding tensor, and the number of characters of the target model, respectively, with
  $K = K'(K'\!+\!1)(\cdots)(K'\!+\!k\!-\!1)/k!$.}\label{tab:ModelExx} 
\end{table}

Markov invariants (see \citet{sumner2008,sumner2009}, and \S \ref{sec:appendix}) are formally polynomials in the components of a phylogenetic tensor $P$ with components $P_{a_1a_2\ldots a_L}$ representing the probability of observing the character state pattern $(a_1,a_2,\ldots,a_L)$ at the leaves of the tree. 
Hence for $L$ taxa $P$ is a tensor with $k^L$ components, indexed by $L$ sets of $k$ multi-indices. Markov invariants are constructed such that, under time evolution associated with the model on the pendant edges, they change at most by a multiplicative factor. 
For clarity, at the pendant edges of the tree let $m_i$, $i=1, \cdots, L$,  be the $K'\times K'$ transition matrices of the starting model, and denote the embedding into the target model\footnote{In the previous discussion the corresponding rate matrices were distinguished by a superscript, $^{(K)}$.} as $M_i \equiv M(m_i)$, $i=1, \cdots, L$. 
Then given the transformation rule, (\ref{eq:PtransfRule}), for $P$ for pendant edge evolution under the model, a Markov invariant $I$ must satisfy (\ref{eq:MarkovInvDefn}), namely
\[
I(P') =  \det(m_1)^{w_1} \det(m_2)^{w_2} \cdots \det(m_L)^{w_L} I(P),
\]
for some integers $w_i$.
We note that for a continuous-time Markov model with rate matrices $Q_i$, we have $m_i=e^{Q_i \tau}$ and $\det(m_i)=e^{ tr(Q_i)\tau}$, as in the introduction.

Recall that a \emph{partition} $\mu$ of an non-negative integer $m$ is a set of non-negative integers $\lambda_1,\lambda_2,\ldots, \lambda_r$ such that $\lambda_1+\lambda_2+\ldots +\lambda_r=m$.
It is usual to write $\mu=(\lambda_1,\lambda_2,\ldots,\lambda_r)$ with $\lambda_1\geq \lambda_2 \geq \ldots \geq \lambda_r$ and to use exponents for repeated parts.
For example we write $\mu=(4,3,3,3,2,2,1)\equiv (4,3^3,2^2,1)$, with $\mu$ being a partition of $18$.
Markov invariants $I$ of degree $D$ are then identified by associating them with certain partitions of special shape, which in turn label certain irreducible group representation characters of the general linear group. 
 
Whether admissible $\mu$ arise at each $D$ and number of leaves $L$, 
and how many occurrences thereof, must be answered for each case by evaluating a certain representation-theoretic branching rule (see \S \ref{sec:appendix}, Theorems 1 \& 2, for details). Instances of such invariants are enumerated in Table \ref{tab:InvtExx}  for the cases identified in Table \ref{tab:ModelExx}. They are listed by $K'$, $k$, $K$ (to define the embedding type), by $L$ for small numbers of leaves, and then by degree $D$ up to 4. From the tables it is evident that there exists a plethora of Markov invariants for embedded submodels. Further information on the independent invariants for phylogenetic tensors constructed under the Markov model can be accessed by studying the isotropy subgroup of leaf permutations on a tree, as in \citet{sumner2009}. 
We defer discussion on the general results, including commentary on cases of possible biological interest, to the conclusions. 

\begin{table}[tbp]
  \centering 
  \begin{tabular}{ccc}
  
  \begin{tabular}[tbp]{|c|c|c|c|}
\hline
 &  $D$  &  $L$ &  \\
  \hline
  \hline
 (2,2,3)  & 1 & 2 & 1 \\
 &  & 3 & 1\\
&  & 4 & 1 \\
\cline{2-4}
  & 2 & 2 & 5 \\
 &  & 3 & 14 \\
&  & 4 & 41\\
\cline{2-4}
& 3 & 2 & 9 \\
&  & 3 & 58 \\
&  & 4 & 401 \\
\cline{2-4}
& 4 & 2 & 23\\
&  & 3 & 321\\
&  & 4 & 5597 \\
\hline \hline
 (2,3,4)  & 1 &1&1 \\
  &   &3&1\\
  &   &4&1\\
\cline{2-4}
 & 2 &2&8\\
  &   &3& 32\\
  &   &4& 128 \\
\cline{2-4}
  & 3 &2&26 \\
  &   &3&292\\
  &   &4&3 464\\
\cline{2-4}
  & 4 &2& 100 \\
  &   &3& 3 688 \\
  &   &4& 158 384 \\
\hline \hline
\end{tabular}

&

\begin{tabular}[tbp]{|c|c|c|c|}
\hline
 &  $D$  &  $L$ &  \\
  \hline
  \hline
(3,2,6)  & 1 & 2 & 1 \\
 &  & 3 & 1\\
&  & 4 & 1 \\
\cline{2-4}
 & 2 & 2 & 0 \\
 &  & 3 & 0 \\
&  & 4 & 0 \\
\cline{2-4}
& 3 & 2 & 2\\
&  & 3 & 4\\
&  & 4 & 8\\
\cline{2-4}
& 4 & 2 & 4 \\
&  & 3 & 31 \\
&  & 4 &  274\\
\hline \hline
 (3,3,10)  & 1 &2&1 \\
  &   &3&1\\
  &   &4&1\\
\cline{2-4}
 & 2 &2& 0\\
  &   &3& 0 \\
  &   &4& 0 \\
\cline{2-4}
  & 3 &2& 5\\
  &   &3&13\\
  &   &4&41\\
\cline{2-4}
  & 4 &2& 19 \\
  &   &3& 338 \\
  &   &4& 6 532 \\
\hline \hline
\end{tabular} &


\begin{tabular}[tbp]{|c|c|c|c|}
\hline
 &  $D$  &  $L$ &  \\
  \hline
  \hline
 (4,2,10)  & 1 &2& 1\\
  &   &3&1\\
  &   &4&1\\
\cline{2-4}
  & 2 &2& 0\\
  &   &3& 0\\
  &   &4& 0\\
\cline{2-4}
  & 3 &2& 0 \\
  &   &3& 0\\
  &   &4& 0\\
\cline{2-4}
  & 4 &2&2 \\
  &   &3&4\\
  &   &4&8\\
\hline \hline
 (4,3,20) & 1 & 2 &1 \\
 &  & 3 &1 \\
&  & 4 &1 \\
\cline{2-4}
 & 2 & 2 & 0\\
 &  & 3 & 0\\
&  & 4 & 0 \\
\cline{2-4}
& 3 & 2 & 0 \\
&  & 3 & 0\\
&  & 4 & 0 \\
\cline{2-4}
& 4 & 2 & 2 \\
&  & 3 & 4 \\
&  & 4 & 8 \\
\hline \hline
\end{tabular}

\end{tabular}
\caption{Enumeration of linearly independent candidate Markov invariants
for the embedded models listed in Table \ref{tab:ModelExx} above, for
small numbers of taxa $L$, and degrees $D$ up to 4. $(K',k,K)$ gives the
number of characters of the progenitor model, the rank of the embedding
tensor, and the number of characters of the target model, respectively.
The linear invariants simply record overall probability conservation for
each phylogenetic tensor. The invariants $I_{3,1}$ and $I_{2,2}$ studied
in this paper are the two nonzero algebraically independent quadratic
invariants from the count of 5 identified for the (2,2,3) model for
$D=L=2$. }\label{tab:InvtExx} 
\end{table}
For now we resume consideration of the lowest dimensional situation which motivated the present study, $2 \hookrightarrow 3$, and the lowest-degree (quadratic) invariants for the simplest situation of two leaves ($L=2$), namely $K'=2$, $k=2$, $K=3$, $L=2$, $D=2$.
Here we outline briefly the manner in which these objects are constructed by standard tensor symmetrisation techniques. The end result will be the explicit forms (\ref{eq:ExplicitI31}) and (\ref{eq:ExplicitI22}) below. 

Recall that we handle the $3$ state embedded model via a rank two phylogenetic probability array $p_{a_1a_2}=p_{a_2a_1}$. 
Given the probability sum 
$p_{11}+ p_{12}+p_{21}+p_{22}=1$, the correct transcription between the two-state and three-state basis is a relabelling $p_{11}\rightarrow P_1$,  $p_{12}\rightarrow 
\frac 12 P_2$, $p_{21}\rightarrow \frac 12 P_2$, $p_{22}\rightarrow P_3$ between $p_{a_1a_2}$ and a three component vector $P_i$. 
By the same token, for a model on 2 leaves, the phylogenetic tensor will be an object $p_{ab, \alpha \beta}$ built from edge transition matrices and a root probability in the usual way. Quantities at quadratic degree  
$p_{ab, \alpha \beta}p_{cd, \gamma\delta}$ therefore admit only certain compatible tensor symmetrisations between the index labels $a,b,c,d$ and $\alpha,\beta,\gamma,\delta$. 
Table
\ref{tab:InvtExx} lists 5 invariants corresponding to symmetry types identified in the discussion in \S \ref{sec:appendix}.
Here we take up the invariants $I_{3,1} $, and $I_{2,2}$, respectively. 

Consider for example the $\lambda=(3,1)$ form quadratic in the components\footnote{Writing $p_{ab,\alpha \beta}$ as $p({ab,\alpha \beta})$ for clarity.}, written down according to standard row and column Young symmetrisation and antisymmetrisation operations (on the sets $a,b,c,d$ and $\alpha,\beta,\gamma,\delta$ separately)\footnote{See \citep{sumner2006a} for details}:
\begin{align}
{\mathcal W}
{\kern -1ex}\raisebox{1ex}{${}_{\!\begin{array}{l}abc\\[-1ex]d\end{array}{\kern -1ex};{\kern -1ex}\begin{array}{l}\alpha \beta \gamma\\[-1ex]\delta \end{array}}$} = & \, 
p(ab,\alpha \beta) p(cd,\gamma \delta) +p(ac,\alpha \beta) p(bd,\gamma \delta) -p(bd,\alpha \beta) p(ac,\gamma \delta) -p(cd,\alpha \beta) p(ab,\gamma \delta) \nonumber \\[-2ex]
& \,+ p(ab,\gamma \beta) p(cd,\alpha \delta) +p(ac,\gamma \beta) p(bd,\alpha \delta) -p(bd,\beta \gamma) p(ac,\alpha \delta) -p(cd,\beta \gamma) p(ab,\alpha \delta) \nonumber
\end{align}
The next step is to identify the part of this array which provides the Markov invariant. This is most natural in a transformed basis for the character states in which the probability mass is treated as a separate (constant) component `0' or `$*$' (for details see Appendix~A of \citet{sumner2008}). In the present case, after implementing this basis transformation, the unique invariant component (identified as $I_{3,1}$) becomes (up to an overall factor):
\[
I_{3,1}:= {\mathcal W}
{\kern -1ex}\raisebox{.5ex}{${}_{\!\begin{array}{l}000\\[-1ex]2\end{array}{\kern -1ex};{\kern -1ex}\begin{array}{l}000\\[-1ex]2 \end{array}}$}
\equiv \sum_{a,b,c = 1}^2 \sum_{\alpha,\beta,\gamma = 1}^2
{\mathcal W}
{\kern -1ex}\raisebox{1ex}{${}_{\!\begin{array}{l}abc\\[-1ex]2\end{array}{\kern -1ex};{\kern -1ex}\begin{array}{l}\alpha \beta \gamma\\[-1ex]2 \end{array}}$}. 
\]
Finally, reverting to the natural 3 state basis $i,j,k = 1,2,3$ interpreting $p_{ab,\alpha \beta}$ as a $3 \times 3$ array $P_{ij}$ via the rules $11\rightarrow 1$, $12 \rightarrow \frac12 2 \leftarrow 21$, $22\rightarrow 3$ already discussed, we have the concrete realization
\begin{align}
\label{eq:ExplicitI31}
I_{3,1} = & \, 4 \big(P_{33} +\textstyle{\frac 12} (P_{23} + P_{32}) + \textstyle{\frac 14} P_{22}\big) -  \nonumber \\
& \, 4 \big(\textstyle{\frac 12} P_{12} + \textstyle{\frac 12} P_{22} + \textstyle{\frac 12} P_{32} +  P_{13} +P_{23} + P_{33}\big)\!\cdot\!
\big(\textstyle{\frac 12} P_{21} + \textstyle{\frac 12} P_{22} +  \textstyle{\frac 12} P_{23}  + P_{31} +P_{32} + P_{33}\big). 
\end{align}
An analogous procedure yields the invariant
\begin{align}
\label{eq:ExplicitI22}
 I_{2,2} = & \,
P_{33} + 2\big(P_{33} +\textstyle{\frac 12} (P_{23} + P_{32}) + \textstyle{\frac 14} P_{22}\big)^2 + \nonumber \\
& \, 
\big(P_{13} +P_{23} + P_{33}\big)\!\cdot\!\big(P_{31} +P_{32} + P_{33}\big) \nonumber \\
& \, 
-2(\textstyle{\frac 12} P_{12} + \textstyle{\frac 12} P_{22} + \textstyle{\frac 12} P_{32} +  P_{13} +P_{23} + P_{33}\big)\!\cdot\!
\big(\textstyle{\frac 12} P_{23} + P_{33}\big) \nonumber \\
& \, -2\big(\textstyle{\frac 12} P_{21} + \textstyle{\frac 12} P_{22} +  \textstyle{\frac 12} P_{23}  + P_{31} +P_{32} + P_{33}\big)\!\cdot\!
\big(\textstyle{\frac 12} P_{32} + P_{33}\big).
\end{align}
The crucial property of the $I_{3,1} $ and $I_{2,2}$ quantities as invariants of weight (from \S \ref{sec:appendix}) $w=1$ and $w=2$, respectively, is how they transform under phylogenetic evolution.
Namely, as $t\rightarrow t+\tau$, we have 
\begin{align}
\label{eq:TransfRulesForI}
I_{3,1}\rightarrow I'_{3,1}  = & \,  \det(m_1) \det(m_2) I_{3,1}, \qquad I_{2,2}\rightarrow I'_{2,2}=  \det(m_1)^2 \det(m_2)^2 I_{2,2},
\end{align}
where $m_{i}=e^{ Q_i\tau}$.

Table \ref{tab:InvtExx} also lists several invariants at degree $3$ (see \S \ref{sec:appendix}). 
One of these, $I_{3,3}$ in our notation, is nothing but the Markov invariant coming from the general Markov model on 3 states, and well known via its log (as used for distance and likelihood studies) as the log Det measure\footnote{See \citet{sumner2008} for comments on the relationship between log Det and other Markov invariants for the general model, and also \citet{sumner2005,sumner2006}.}, whose form (as a cubic polynomial) is obvious and will provide a standard by which the behaviour of $I_{3,1} $ and $I_{2,2}$ under simulation can be compared (see the concluding results and discussion below):
\[
\mbox{Det} \equiv I_{3,3} =
 P_{11}P_{22}P_{33} +  P_{12}P_{23}P_{31} +  P_{13}P_{32}P_{21}  
  - P_{11}P_{23}P_{32} -  P_{13}P_{22}P_{31} -  P_{12}P_{21}P_{33}. 
\]
Of course the $\mbox{Det}$ is itself a Markov invariant of weight $w=3$ and satisfies the transformation rule
\beqn
\mbox{Det}\rightarrow \mbox{Det}'=\det(m_1)^3 \det(m_2)^3\mbox{Det}.\nonumber
\eqn

%% file: sec4.tex
\section{Results}
\label{sec:Results} 
\noindent
Is this section we present a simple simulation study that compares the performance of the Markov invariants $\mbox{Det}$, $I_{22}$ and $I_{31}$ as pairwise distance estimators for data generated under our symmetric embedded model.
We do this by taking the theoretical probability distribution for a two leaf tree generated under the model and then sampling from the multinomial distribution for a range of sequence lengths.
We discuss the derivation of \emph{unbiased} estimators of the invariants and observe that the invariants $I_{22}$ and $I_{31}$ have superior statistical estimation power over that of the (log) $\mbox{Det}$.
This adds credibility to the intuitive notion that it should be invariants of lower degree and lower weight that can be expected to perform best in practical contexts.

If we take a root probability distribution $\pi_i$, $i=1,2,3$, the ``starting'' (zero-edge length) probability distribution on a two leaf tree is $P^\circ_{ii}= \pi_i$, $P^\circ_{ij}=0$ if $i\ne j$. 
For non-zero edge lengths, $I_{3,1}$ and $I_{2,2}$ are then determined by this $P^\circ$ and the transformation rules (\ref{eq:TransfRulesForI}).
Starting with the rate matrix $Q$ as in (\ref{eq:Qmatrix2by2}), we have the standard form
\[
m(t) = e^{tQ} = \left( \begin{array}{cc} 1\!-\!\alpha \lambda & \beta \lambda \\
                           \alpha \lambda & 1\!-\!\beta  \lambda \end{array}\right),
\]
where $\lambda = \frac{1}{\alpha+\beta}{(1-e^{-(\alpha\!+\!\beta)t})}$, with determinant $\det m =
\exp(-(\alpha + \beta)t)$. 
Thus choosing as independent parameters $\alpha$ and $t$, with $\alpha + \beta = 1$, if the edge distances are $t_1$ and $t_2$, the theoretical values for $I_{3,1}$ and $I_{2,2}$ become after evaluating them on $P^\circ$,
\[
I_{3,1} = e^{-t_1}e^{-t_2} (4\pi_1\pi_3 + \pi_1\pi_2 + \pi_2\pi_3),
\qquad I_{2,2} = e^{-2t_1}e^{-2t_2}\textstyle{ \frac 18}(\pi_2^2+8 \pi_1\pi_3).
\]
These are to be compared to the ``Det'' function, with theoretical value 
\[
Det = e^{-3t_1}e^{-3t_2} (\pi_1 \pi_2\pi_3).
\]

We take as data an alignment of two sequences consisting of three character states corresponding to $i=1,2,3$ reduced to the pattern frequencies
\[F_{ij}:=\{\textit{number of occurrences of the pattern }(ij)\}.\] 
It is possible to use this data and the above invariant functions to obtain an estimator of the pairwise distance $\Delta=t_1+t_2$ under the symmetric embedded model $2\hookrightarrow 3$, as follows.

Consider the estimators of the invariants constructed by simply making the ``obvious'' replacement $P_{ij}\rightarrow f_{ij}:=\textstyle{\frac{1}{N}}F_{ij}$ given by\footnote{More formally, this equates to taking the $f_{ij}$ as our best estimate of the probabilities $P_{ij}$ for this data.} 
\beqn
\widehat{I}_{3,1} := & \, 4 \big(f_{33} +\textstyle{\frac 12} (f_{23} + f_{32}) + \textstyle{\frac 14} f_{22}\big) -  \nonumber \\
&  \, 4 \big(\textstyle{\frac 12} f_{12} + \textstyle{\frac 12} f_{22} + \textstyle{\frac 12} f_{32} +  f_{13} +f_{23} + f_{33}\big)\!\cdot\!
\big(\textstyle{\frac 12} f_{21} + \textstyle{\frac 12} f_{22} +  \textstyle{\frac 12} f_{23}  + f_{31} +f_{32} + f_{33}\big),
\\
\widehat{I}_{2,2} := & \,
f_{33} + 2\big(f_{33} +\textstyle{\frac 12} (f_{23} + f_{32}) + \textstyle{\frac 14} f_{22}\big)^2 + 
\big(f_{13} +f_{23} + f_{33}\big)\!\cdot\!\big(f_{31} +f_{32} + f_{33}\big) \nonumber \\
& \, 
-2(\textstyle{\frac 12} f_{12} + \textstyle{\frac 12} f_{22} + \textstyle{\frac 12} f_{32} +  f_{13} +f_{23} + f_{33}\big)\!\cdot\!
\big(\textstyle{\frac 12} f_{23} + f_{33}\big) \nonumber \\
& \, -2\big(\textstyle{\frac 12} f_{21} + \textstyle{\frac 12} f_{22} +  \textstyle{\frac 12} f_{23}  + f_{31} +f_{32} + f_{33}\big)\!\cdot\!
\big(\textstyle{\frac 12} f_{32} + f_{33}\big),
\\
\widehat{Det}:=& f_{11}f_{22}f_{33} +  f_{12}f_{23}f_{31} +  f_{13}f_{32}f_{21}  
  - f_{11}f_{23}f_{32} -  f_{13}f_{22}f_{31} -  f_{12}f_{21}f_{33}.
\eqn
We work under the assumption that frequency array $\left[F_{ij}\right]_{1\leq i,j \leq 3}$ is generated by sampling $N$ patterns with probability $p_{ij}$ generated under our model on a two taxa tree. 
This means that the probability of observing a given frequency array $\left[F_{ij}\right]_{1\leq i,j \leq 3}$ is given by the multinomial form
\beqn
\frac{N!}{F_{11}!F_{12}!\ldots F_{33}!}P_{11}^{F_{11}}P_{12}^{F_{12}}\ldots P_{33}^{F_{33}}.\nonumber
\eqn
Under these conditions it is easy to show using generating function techniques that we have
\beqn
E\left[\widehat{I}_{3,1}\right]=N(N-1)I^{(2)}_{3,1}+N(I^{(1)}_{3,1}-(P_{22}+2P_{32}+2P_{23}+4P_{33})),\nonumber
\eqn
where $I^{(2)}_{3,1}$ and $I^{(1)}_{3,1}$ are the quadratic and linear parts of $I_{3,1}$ respectively (see \citet{sumner2008} for example calculations of this kind).
Thus we observe that $E\left[\widehat{I}_{3,1}\right]\neq I_{3,1}$, so that $\widehat{I}_{3,1}$ is a \emph{biased} estimator of $I_{3,1}$. 

This is easily rectified by defining the \emph{unbiased} estimator:
\beqn
\widehat{I}_{3,1}^{(ub)}:=\frac{1}{N(N-1)}\left(\widehat{I}^{(2)}_{3,1}+(N-1)\widehat{I}^{(1)}_{3,1}+F_{22}+2F_{22}+2F_{23}+2F_{32}+4F_{33}\right).\nonumber
\eqn 
This estimator then takes on the expectation value
\beqn
E\left[\widehat{I}_{3,1}^{(ub)}\right]=I_{3,1}=e^{-(t_1+t_2)}(4\pi_1\pi_3 + \pi_1\pi_2 + \pi_2\pi_3).\nonumber
\eqn

A similar observation for $I_{2,2}$ leads to its unbiased estimators
\beqn
\widehat{I}_{2,2}^{(ub)}&:=\textstyle{\frac{1}{N(N-1)}}\left(\widehat{I}^{(2)}_{2,2}+(N-1)\widehat{I}^{(1)}_{2,2}+(F_{33}-\textstyle{\frac 18}F_{22}+\textstyle{\frac 12}F_{23}+\textstyle{\frac 12}F_{32})\right),\nonumber\\
\widehat{\mbox{Det}}^{(ub)}&:=\textstyle{\frac{1}{N(N-1)(N-2)}}\widehat{\mbox{Det}},
\eqn
with expectation values
\beqn
E\left[\widehat{I}_{2,2}^{(ub)}\right]&=I_{2,2}=e^{-2(t_1+t_2)}\textstyle{\frac 18}(\pi_2^2 + 8\pi_1\pi_3),\nonumber\\
E\left[\widehat{\mbox{Det}}^{(ub)}\right]&=Det=e^{-3(t_1+t_2)}(\pi_1\pi_2\pi_3).
\eqn 

We now define the marginalizations $F^{(1)}_i:=\sum_j F_{ij}$ and $F^{(2)}_j:=\sum_i F_{ij}$ and, as is standard for the ``log Det'', estimate the values $\pi_1,\pi_2,\pi_3$ by assuming the process was stationary and taking the harmonic means:
\[
\widehat{\pi_i}=\sqrt{F^{(1)}_iF^{(2)}_i}.
\]

With these estimators in hand, we are now in a position to define six reasonable estimators of $\Delta=t_1+t_2$ that can readily be evaluated directly from the pattern counts $F_{ij}$:
\beqn\label{eq:distestimates}
\Delta_{3,1}&:=-\log\left(\widehat{I}_{3,1}\right)+\log\left({4\hat{\pi}_1\hat{\pi}_3+\hat{\pi_1}\hat{\pi_2}+\hat{\pi}_2\hat{\pi}_3}\right),\\
\Delta_{2,1}&:=-\log\left(\widehat{I}_{2,1}\right)+\log\left({\hat{\pi_1}^2+8\hat{\pi_1}\hat{\pi_3}}\right),\\
\Delta_{\mbox{Det}}&:=-\log\left(\widehat{Det}\right)+\log\left(\hat{\pi_1}\right)+\log\left(\hat{\pi_2}\right)+\log\left(\hat{\pi_3}\right),\\
\Delta_{3,1}^{(ub)}&:=-\log\left(\widehat{I}_{3,1}\right)+\log\left({4\hat{\pi}_1\hat{\pi}_3+\hat{\pi_1}\hat{\pi_2}+\hat{\pi}_2\hat{\pi}_3}\right),\\
\Delta_{2,1}^{(ub)}&:=-\log\left(\widehat{I}_{2,1}\right)+\log\left({\hat{\pi_1}^2+8\hat{\pi_1}\hat{\pi_3}}\right),\\
\Delta_{\mbox{Det}}^{(ub)}&:=-\log\left(\widehat{Det}^{(ub)}\right)+\log\left(\hat{\pi_1}\right)+\log\left(\hat{\pi_2}\right)+\log\left(\hat{\pi_3}\right).
\eqn

\begin{figure}[htb]
\caption{\textbf{Comparison of performance of Markov invariants.}  The barplots give the mean (dark shade) and variance (light shade) of $10^7$ runs with sequence length ranging from $N=100$ to $N=100,000$.}
\vspace{1em}
\centering
\scalebox{.75}{\includegraphics{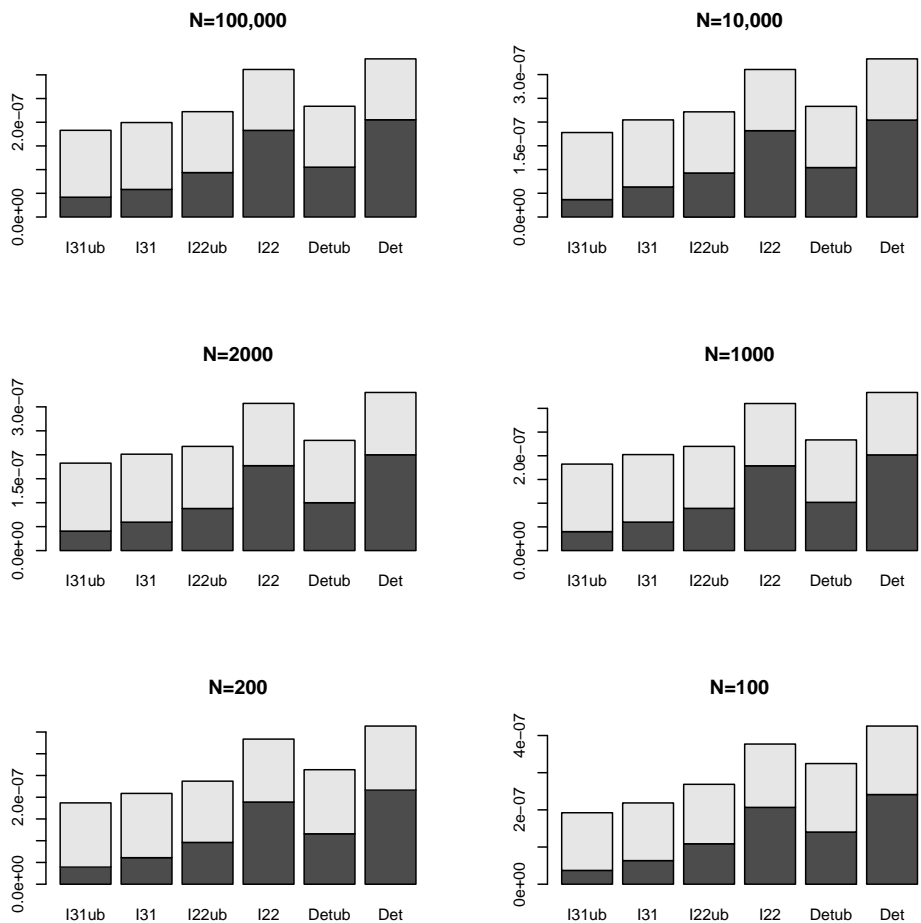}}
\label{fig:sim1}
\end{figure}

To compare the performance of these estimators we performed a simulation study over a range of sequence lengths, from very long $N=10^6$ to very short $N=10$, with fixed rate parameters $\alpha=0.45,\beta=0.55$ and $t_1+t_2=1$.
The results are presented in Figure~\ref{fig:sim1}. Careful inspection of the results shows that it is consistently the lower degree invariants that have greater statistical power and that taking unbiased forms also provides a significant improvement.

%% file: conc.tex
\section{Conclusion}
In this paper we have described a novel approach to phylogenetic model construction using symmetric tensor embeddings that ensures multiplicative closure.
Although this model construction is of interest in its own right, we went further to exploit the simple structure of the embedding to give examples of Markov invariants for these models.
We also showed how these invariants can be exploited effectively as distance estimators with favourable statistical properties (as compared to the standard ``log Det'').

We should emphasize that we do not make any claim that the symmetrically embedded models discussed have any particular direct appeal as biologically realistic rate matrices for molecular phylogenetics -- this is indeed why we did not shy away from considering the 3-state case in detail.
We do however argue strongly that the multiplicative closure that is present in our model is itself highly desirable from a biological perspective, and make the point that the general time reversible model (perhaps the most popular currently in use) does not satisfy the closure property. 

The work presented in this paper thus serves as an elementary example that illustrates how one may go about constructing models with multiplicative closure.
We are currently directing work into expanding our knowledge of such ``closed'' phylogenetic models and we expect that furthering the connection to Lie algebras will be vital in this regard.

%% file: appendix.tex
\begin{appendix}
\setcounter{equation}{0}
\renewcommand{\theequation}{{A}-\arabic{equation}}
\section{Appendix: Groups and representation theory in phylogenetic models}
\label{sec:appendix}
\noindent
\subsection{General results}
The Markov model for phylogenetic branching sketched in \S \ref{sec:SymmEmbedd} above can be given a formal setting for considering phylogenetic invariants and related constructions, and in particular, Markov invariants \citep{sumner2008}.  In this appendix we recall our main results from that paper, and the representation-theoretic results which, appropriately generalised here, lead to the enumeration and construction of Markov invariants for the embedded submodels which are the subject of the present work.

The starting point of our approach is to regard the model of stochastic change via Markov matrices, (\ref{eq:LinearAction}), in terms of linear actions of certain matrix groups $G$ affiliated with stochastic matrices \citep{johnson1985,mourad2004}. Most generally, consider a phylogenetic tensor $P$ of rank $L$ (the array of pattern frequencies for $L$ leaf edges of a presumed phylogenetic tree under the general Markov model). The equation equivalent to (\ref{eq:LinearAction}) describing the evolution along the leaf edges\footnote{
Here we do not need to consider the full phylogenetic tree model, which involves labelling all edges by stochastic matrices and appropriate summations over characters. For remarks on such internal structure in the context of Markov invariants see \citep{sumner2008,sumner2009}.},
\begin{align}
\label{eq:PtransfRule}
P' = & \, M_1 \otimes M_2 \ldots \otimes M_L \cdot P ,
\end{align}
is simply the action of an element $(M_1 , M_2, \cdots, M_L)$ of the $L$-fold direct product group $G \times G \times \cdots \times G$. 
Where there is a rate matrix in continuous time models, or generically if the $M_i$ are nonsingular, then $G$ can simply be taken to be the general linear group $GL(K)$. Specifically however, we restrict consideration to probability-conserving, stochastic matrices fulfilling a unit column\footnote{Or unit row sum condition; however we use left multiplication for our actions.} sum condition, so the relevant matrix group is $GL_1(K)$, which from
\citep{johnson1985,mourad2004} is isomorphic to the affine group\footnote{We discuss groups and representions over $\mathbb{C}$ and regard real parametrisations as a separate issue.} $\cong GL(K-1) \ltimes \mathbb C^{K\!-\!1}$.

Including the symmetric embeddings construction, the full chain of edge subgroups becomes
\begin{align}
\label{eq:SubgroupChain}
GL(K) > & \, GL_1(K) 
> GL_1(2) > GL(1) .
\end{align}
The last step simply asserts that with a fixed rate matrix $Q$, a continuous-time Markov chain generically provide a $K \times K$ representation of the time evolution group ${\mathbb R}^+$; the specialisation $GL_1(K) > GL_1(2)$ describes our present `$2\hookrightarrow K$' situation, and would be replaced for example, by 
$GL_1(K) 
> GL_1(K')$ in the `$K'\hookrightarrow K$' case, $K'<K$. Formally then, the assertion (\ref{eq:PtransfRule}) is that of the full group branching rule for the reduction
\begin{align}
\label{eq:FullSubgroupChain}
GL(K^L) > & \, GL(K) \times GL(K) \times \cdots \times GL(K) 
\end{align}
followed by the reduction (\ref{eq:SubgroupChain}) on each edge.  From the formal point of view, the problem of dealing with polynomial functions of $P$, say of degree $D$, is thus to implement the corresponding branching rule for the group representations arising. Specifically the \emph{Markov invariants} $I(P)$, now for the embedded $GL_1(2)$ submodel, or $GL_1(K')$ in general, are in correspondence with the 1-dimensional
representations occurring. Specifically, representing $M_i = M(m_i)$, with $m_i \in GL_1(K')$, $i=1,\cdots,L$ as the embedded $K'\hookrightarrow K$ submodel, then under (\ref{eq:PtransfRule}) we have by definition
\begin{align}
\label{eq:MarkovInvDefn}
I(P') = & \, \det(m_1)^{w_1} \det(m_2)^{w_2} \cdots \det(m_L)^{w_L} I(P) 
\end{align}
for a Markov invariant of weight $(w_1,w_2,\cdots,w_L)$.

The two stages of the above representation-theoretic problem have been solved in \cite{sumner2008}, and for completeness we quote the relevant theorems. Firstly recall that the polynomial ring ${\mathbb C}{[}P{]}$ is isomorphic to the symmetric tensor algebra $\vee(P)$, that is, at each degree, symmetric tensor powers of the module corresponding to $P$. 

\noindent
\textbf{Theorem 1: Polynomial covariants for embedded models.} \\
Consider the embedding $GL(K^L) \supset  \times^L GL(K')$ defined by the branching rule for the fundamental $K^L$-dimensional representation
\[
\{1\} \rightarrow  \{\lambda_1\} \otimes  \{\lambda_2\} \otimes \cdots \otimes  \{\lambda_L\}.
\]
The corresponding branching rule for the $D$'th symmetric tensor power is given by
\begin{align}
\label{eq:LeafBranchingRule}
\{ D \} \longrightarrow & \, \sum_{\sigma_i \vdash D}^{\sigma_1 * \sigma_2 * \cdots * \sigma_L \ni (D)}
\big(\{\lambda_1 \} \underline {\otimes} \{ \sigma_1 \}\big) \otimes \big(\{\lambda_2 \} \underline {\otimes} \{ \sigma_2 \}\big) \otimes \cdots \otimes \big(\{\lambda_L \} \underline {\otimes} \{ \sigma_L \}\big) .
\end{align}
Here standard partition notation $\lambda$ or $( \cdot )$ has been adopted for irreducible tensor representations, with $\{ \cdot \}$ denoting the corresponding characters (symmetric functions, as in \citep{littlewood1955}).
The operation $*$ of \emph{inner multiplication} corresponds to the evaluation of tensor products of irreducible representations in the symmetric group ${\mathfrak S}_D$ for partitions $\sigma_i \vdash D$. The symbol
$\underline{\otimes}$ stands for the operation of \emph{plethysm} on group characters. For a recent discussion of the calculus of plethysms see \citep{fauser2006}.

\noindent
The above result has been stated in full generality allowing for the possibility of \emph{heterogeneous} edges (models with different numbers of characters). In practice we restrict attention to standard embeddings $K'\hookrightarrow K$ where $K$ is the dimension of a certain irreducible representation $\lambda$ of $GL(K')$. Moreover, for the cases of interest,
$\lambda$ is a symmetric tensor of rank $k$; specifically for $2\hookrightarrow K$, $K=k+1$. With the above branching rule in hand, the occurrence of one dimensional representations (including multiplicities) can be read off using the following result:\\

\noindent
\textbf{Theorem 2: Polynomial invariants for phylogenetic models and embedded models.} \\
Linearly independent polynomial invariants at degree $D$ of the groups 
(i) $\times^L GL(K')$, (ii) $\times^LGL_1(K')$ and (iii) $\times^L GL_{1,1}(K')$ for the general phylogenetic model are given by the one dimensional modules of these groups within polynomial representations of $\times^L GL(K')$  corresponding to the following partitions:
\begin{align}
\times^L GL(K'): & \, \qquad \{r^{K'}\} \otimes \{r^{K'}\}\otimes \cdots \otimes \{r^{K'}\},  \quad \mbox{where} \quad  {K'} r=kD, \nonumber  \\
\times^L GL_1(K'): & \, \qquad \{r_1\!+\!s_1,r_1^{K'\!-\!1}\} \otimes \{r_2\!+\!s_2,r_2^{K'\!-\!1}\}\otimes \cdots \otimes \{r_L\!+\!s_L,r_L^{K'\!-\!1}\}, \\ 
& \, \qquad  \mbox{where} \quad K' r_i +s_i =kD, \nonumber  \\
\times^L GL_{1,1}(K'): & \, \qquad \{r_1\!+\!s_1,r_1^{K'\!-\!2},t_1\} \otimes \{r_2\!+\!s_2,r_2^{K'\!-\!2},t_2\}\otimes \cdots \otimes \{r_L\!+\!s_L,r_L^{K'\!-\!2},t_L\},  \nonumber \\
& \, \qquad  \mbox{where} \quad (K'\!-\!1) r_i +s_i+t_i =kD, \nonumber 
\end{align}
The number of admissible partitions of the given forms $\{\mu_1\} \otimes \{\mu_2\} \otimes \cdots \otimes \{\mu_L \}$ in each case (i), (ii), (iii), deriving from (\ref{eq:LeafBranchingRule}), is the number of times the inner product
$\sigma_1 *\sigma_2 * \cdots * \sigma_L$ of irreducible representations of the symmetric group ${\mathfrak S}_D$ contains the one-dimensional irreducible representation $(D)$. This is also the number of linearly independent polynomial invariants in each case. \hfill $\Box$

\noindent
Clearly the case $k=1, K'=K$ corresponds to the standard situation treated in \citep{sumner2008}, whereas $k\ne1, K'<K$ covers embedded submodels. For the $K'=2$ model (general phylogenetic model on 2 characters) and polynomial degree $D$, the Markov invariants (invariants of $GL_1(2)$, and \emph{a fortiori} of $GL(2)$) occur as characters of the type $\{r,r\}$ with $ 2r = kD$ (one dimensional tensor representations of $GL(2)$), denoted $I_{r,r}$, of weight $r$ or more generally are associated with characters of the type $\{r+s,r\}$ with $2r+s = kD$, denoted $I_{r+s,r}$ also of weight $r$. 

\subsection{Enumeration of invariants for low-dimensional cases}
Finally we turn to the enumeration of invariants for concrete, low-dimensional, cases. 
For the $2 \hookrightarrow 3$, $k=2$, $K'=2$, $K=3$ case, admissible characters (partition shapes) $\{\mu_i\}$ according Theorem 2  are obviously $\{ 4\}$, $\{ 2,2\}$ and $\{3,1\}$ at degree $D=2$, and $\{ 6\}$,$\{ 5,1\}$ and $\{ 4,2\}$ and $\{ 3,3\}$ at degree $D=3$. Their precise occurrence is determined by the evaluations of the appropriate plethysms. The following are well known and can be checked on dimensional grounds:
\begin{align}
\mbox{$D=2$}:&\, \nonumber\\
\{2\} \underline{\otimes} \{2\} = &\, \{4\} + \{2,2\}, \quad \{2\} \underline{\otimes} \{1^2\} = \{3,1\} ;\nonumber \\
\mbox{$D=3$ }:&\, \nonumber\\
\{2\} \underline{\otimes} \{3\} = &\,\{6\} + \{4,2\} +  \{2^3\}, \quad \{2\} \underline{\otimes} \{2,1\} = \{5,1\} + \{4,2\} +\{3^2\},  
\quad \{2\} \underline{\otimes} \{1^3\} = \{41^2\} + \{3,3\}. \nonumber 
\end{align}
In turn, for $L=2$ taxa, the possible invariants are associated with pairs $\{\mu_1\}\otimes \{\mu_2\}$ for such admissible characters, weighted by a combinatorial factor (2 if $\{\mu_1\}\ne \{\mu_2\}$), and weighted by the (nonzero) multiplicity of the trivial representation $(D)$ of the symmetric group ${\mathfrak S}_D$ in the reduction of the inner product of the parent $\{\sigma_1\}* \{\sigma_2\}$. Given the rules 
\begin{align}
\mbox{$D=2$ (in ${\mathfrak S}_2$)}:&\, \nonumber\\
\{2\} * \{2\} = &\, \{2\}, \quad \{2\}* \{1^2\} = \{1^2\}, \quad \{1^2\} * \{1^2\} = \{2\} \nonumber \\
\mbox{$D=3$ (in ${\mathfrak S}_3$)}:&\, \nonumber\\
\{3\} * \{3\} = &\, \{3\}, \quad \{3\} * \{2,1\} = \{2,1\}, \quad\{2,1\}* \{2,1\} = \{3\} + \{1^3\} + \{2,1\},
\nonumber \\
 \{2,1\} * \{1^3\} = & \,   \{2,1\}, \quad \{1^3\} * \{1^3\} = \{3\}, \nonumber 
\end{align}
it is clear that there are $5= 2^2 +1^2$ linearly independent invariants at quadratic degree (namely $\{ 4\} \otimes \{4\}$, $\{3,1\}\otimes\{4\}$,
$\{4\} \otimes \{3,1\}$, $\{3,1\}\otimes\{3,1\}$, and $\{2,2\}\otimes\{2,2\}$), and $9 = 2^2+2^2+1^2$ linearly independent invariants at cubic degree. At the quadratic level,  
$\{ 4\} \otimes \{4\}$ just represents probability conservation for $P$, while $\{3,1\}\otimes\{4\}$ and $\{4\} \otimes \{3,1\}$ turn out to vanish identically for this case\footnote{They are examples of \emph{mixed} weight invariants which in principle give \emph{different} information for each edge (other examples of such mixed invariants were noted in \citep{sumner2008}).}. 
These results verify that at quadratic degree, both $\{2,2\}\otimes \{2,2\}$ and $\{3,1\}\otimes \{3,1\}$ occur in the appropriate plethysms for admissible inner products, and lead to the invariants $I_{2,2}$ and $I_{3,1}$ constructed explicitly in \S \ref{sec:MarkovInv}. Their
properties are explored numerically in the final \S \ref{sec:Results}, via some simple simulation studies. 
At cubic level it turns out that there are various invariants involving $\{4,2\}$, $\{5,1\}$ and \emph{two} invariants $\{3,3\}\otimes\{3,3\}$ (again, cases such as $\{6\}\otimes \{6\}$ are trivial by probability conservation\footnote{In relation to the remarks about algebraic independence below, note that the product $I_4 I_{2,2}$ of the linear invariant $I_{4}$ and the quadratic $I_{2,2}$ is also of cubic degree and has weight $w=2$.}). One of the latter, associated with the ($k=1$, $K'=K=3$) $GL(3)$ antisymmetric invariant, via $\{1^3\} * \{1^3\} = \{3\}$, is in fact identical to the determinant function.  

The above results for $2 \hookrightarrow 3$ prove to be the simplest of a plethora of cases, some of which may be of interest for phylogenetic applications (see the concluding remarks, \S \ref{sec:Results}), but whose existence serves to illustrate our general philosophy. In \S \ref{sec:MarkovInv} we record  invariants for symmetric embeddings for diverse (low-dimensional) cases of initial and target models $K'$, $k$ and $K$ (see table \ref{tab:ModelExx} for a tabulation of models and table \ref{tab:InvtExx} for an enumeration of invariants for them\footnote{At least up to linear independence: formally they form a ring, but the question of algebraic independence is beyond the scope of the present investigation (see \citet{sumner2008}).}). All manipulations with products, plethysms and group branching rules can be evaluated symbolically using an appropriate group theory package. The program \texttt{Schur}, \citep{schur}, works with combinatorial algorithms based on manipulations of the group characters encoded as the celebrated Schur functions (symmetric polynomials in $n$ indeterminates representing the eigenvalues of a $n \times n$ matrix). 

The general algorithm for identification of higher invariants follows the above pattern. Consider for example the following plethysm\footnote{Using \texttt{Schur}.} at degree 5 for a rank 2 embedding:
\begin{align}
\{2\}\otimes\{3,2\} = & \, \{82\} + \{73\} + \{721\} + \{64\} + \{631\} + 2\{62^2 \} + \{541\} + \{532\} +
\nonumber \\
& \, + \{531^2 \}
       + \{52^2 1\} + \{4^2 2\} + \{4321\} + \{42^3 \} + \{3^2 21^2 \}.
\nonumber
\end{align}
Thus admissible $\{\mu\}$ for the 2 state model $2\hookrightarrow 3$ are $\{82\}$,$\{73\}$, and $\{64\}$; for the 3 state model\footnote{According to Theorem 2, for choices of parameters giving a \textit{symmetric} model on 3 states, there would be additional candidates $\{721\}$, $ \{631\}$, $ \{541\}$, $\{532\}$, and $\{4^2 2\}$.} $3\hookrightarrow 6$, we have $ \{6,2^2\}$ with multiplicity 2, and for the 4 state $4\hookrightarrow 10$ model, the candidate\footnote{Likewise for a starting {symmetric} 4 state model, there is the additional candidate $\{5 2^2 1\}$.} $\{42^3 \}$. Enumeration of invariants entails counting those products of admissible $\{\mu \}$ whose parent inner product $\prod_i * \{\sigma_i\}$, weighted by the correct multinomial factor,
of symmetric group characters in ${\mathfrak S}_D$ contains the trivial one-dimensional representation $(D)$ (weighted by multiplicity $\ge 1$).
\end{appendix}

%% file: main.bbl
\begin{thebibliography}{21}
\expandafter\ifx\csname natexlab\endcsname\relax\def\natexlab#1{#1}\fi
\expandafter\ifx\csname url\endcsname\relax
  \def\url#1{\texttt{#1}}\fi
\expandafter\ifx\csname urlprefix\endcsname\relax\def\urlprefix{URL }\fi
\providecommand{\selectlanguage}[1]{\relax}

\bibitem[{Allman \& Rhodes(2009)}]{allman2009a}
\textsc{Allman, E.~A. \& Rhodes, J.~A.} (2009) Private communication.

\bibitem[{Bandelt \& Dress(1992)}]{bandelt1992}
\textsc{Bandelt, H.~J. \& Dress, A. W.~M.} (1992).
\newblock Split {D}ecomposition: A new and useful approach to phylogenetic
  analysis of distance data.
\newblock \emph{Mol. Phylogenet. Evol.} \textbf{1}, 242--252.

\bibitem[{Cavender \& Felsenstein(1987)}]{cavender1987}
\textsc{Cavender, J.~A. \& Felsenstein, J.} (1987).
\newblock Invariants of phylogenies in a simple case with discrete states.
\newblock \emph{J. Class.} \textbf{4}, 57--71.

\bibitem[{Fauser \emph{et~al.}(2006)Fauser, Jarvis, King \&
  Wybourne}]{fauser2006}
\textsc{Fauser, B., Jarvis, P.~D., King, R.~C. \& Wybourne, B.~G.} (2006).
\newblock New branching rules induced by plethysm.
\newblock \emph{J. Phys. A Math. Gen.} \textbf{39}, 2611--2655.

\bibitem[{Holland \& Moulton(2004)}]{holland2004}
\textsc{Holland, B. \& Moulton, V.} (2004).
\newblock \emph{Algorithms in Bioinformatics}, chap. Consensus networks: {A}
  method for visualising incompatibilities in collections of trees.
\newblock Springer, pp. 165--176.

\bibitem[{Huelsenbeck \emph{et~al.}(2004)Huelsenbeck, Larget \&
  Alfaro}]{huelsenbeck2004}
\textsc{Huelsenbeck, J.~P., Larget, B. \& Alfaro, M.~E.} (2004).
\newblock Bayesian phylogenetic model selection using reversible jump {Markov}
  chain {Monte Carlo}.
\newblock \emph{Mol. Biol. Evol.} \textbf{21}, 1123--1133.

\bibitem[{Isaev(2004)}]{isaev2004}
\textsc{Isaev, A.} (2004).
\newblock \emph{Introduction to Mathematical Methods in Bioinformatics}.
\newblock Springer.

\bibitem[{Johnson(1985)}]{johnson1985}
\textsc{Johnson, J.~E.} (1985).
\newblock {M}arkov-type {Lie} groups in {$GL(n,{R})$}.
\newblock \emph{J. Math. Phys.} \textbf{26}, 252--257.

\bibitem[{Lake(1987)}]{lake1987}
\textsc{Lake, J.~A.} (1987).
\newblock A rate-independent technique for analysis of nucleic acid sequences:
  evolutionary parsimony.
\newblock \emph{Mol. Biol. Evol.} \textbf{4}, 167--191.

\bibitem[{Littlewood(1955)}]{littlewood1955}
\textsc{Littlewood, D.~E.} (1955).
\newblock The kronecker product of symmetric group representations.
\newblock \emph{J. Lond. Math. Soc.} \textbf{s1-31(1)}, 89--93.

\bibitem[{Mourad(2004)}]{mourad2004}
\textsc{Mourad, B.} (2004).
\newblock On a {Lie}-theoretic approach to generalised doubly stochastic
  matrices and applications.
\newblock \emph{Linear and Multilinear algebra} \textbf{52}, 99--113.

\bibitem[{Pagel \& Meade(2004)}]{pagel2004}
\textsc{Pagel, M. \& Meade, A.} (2004).
\newblock A phylogenetic mixture model for detecting pattern-heterogeneity in
  gene sequence or character-state data.
\newblock \emph{Syst. Biol.} \textbf{53}, 571--581.

\bibitem[{Posada \& Crandall(1998)}]{posada1998}
\textsc{Posada, D. \& Crandall, K.~A.} (1998).
\newblock Modeltest: testing the model of {DNA} substitution.
\newblock \emph{Bioinformatics} \textbf{14}, 817--818.

\bibitem[{Semple \& Steel(2003)}]{semple2003}
\textsc{Semple, C. \& Steel, M.} (2003).
\newblock \emph{Phylogenetics}.
\newblock Oxford Press.

\bibitem[{Sumner(2006)}]{sumner2006a}
\textsc{Sumner, J.~G.} (2006).
\newblock {Entanglement, Invariants, and Phylogenetics}.
\newblock \emph{PhD thesis, University of Tasmania,
  \texttt{http://eprints.utas.edu.au}} .

\bibitem[{Sumner \emph{et~al.}(2008)Sumner, Charleston, Jermiin \&
  Jarvis}]{sumner2008}
\textsc{Sumner, J.~G., Charleston, M.~A., Jermiin, L.~S. \& Jarvis, P.~D.}
  (2008).
\newblock Markov invariants, plethyms and phylogenetics.
\newblock \emph{J. Theor. Biol.} \textbf{253}, 601--615.

\bibitem[{Sumner \& Jarvis(2005)}]{sumner2005}
\textsc{Sumner, J.~G. \& Jarvis, P.~D.} (2005).
\newblock Entanglement invariants and phylogenetic branching.
\newblock \emph{J. Math. Biol.} \textbf{51}, 18--36.

\bibitem[{Sumner \& Jarvis(2006)}]{sumner2006}
\textsc{Sumner, J.~G. \& Jarvis, P.~D.} (2006).
\newblock Using the tangle: A consistent construction of phylogenetic distance
  matrices.
\newblock \emph{Math. Biosci.} \textbf{204}, 49--67.

\bibitem[{Sumner \& Jarvis(2009)}]{sumner2009}
\textsc{Sumner, J.~G. \& Jarvis, P.~D.} (2009).
\newblock Markov invariants and the isotropy subgroup of a quartet tree.
\newblock \emph{J. Theor. Biol.} \textbf{258}, 302--310.

\bibitem[{Woodhams \emph{et~al.}(2009)Woodhams, Sumner \&
  Charleston}]{woodhams2009}
\textsc{Woodhams, M., Sumner, J.~G. \& Charleston, M.~A.} (2009).
\newblock Mosiac models for phylogenetic estimation.
\newblock \emph{\textit{in preparation}} .

\bibitem[{Wybourne(2004)}]{schur}
\textsc{Wybourne, B.~G.} (2004).
\newblock \texttt{Schur}: An interactive programme for calculating properties
  of {Lie} groups. version 6.03.
\newblock \emph{\texttt{http://sourceforge.net/projects/schur}} .

\end{thebibliography}
